\documentclass[10pt,conference]{IEEEtran}
\IEEEoverridecommandlockouts

\usepackage{subfigure}
\usepackage{colortbl}
\usepackage{bm}
\usepackage{geometry}
\usepackage{graphicx}  % Written by David Carlisle and Sebastian Rahtz
\usepackage{url}       % Written by Donald Arseneau

\usepackage{amsmath}    
\allowdisplaybreaks[4]
\usepackage{cite}

\usepackage{amsfonts,amssymb}

\usepackage{stfloats}
\usepackage{empheq}
\usepackage{cases}
\usepackage{algorithm}
\usepackage{multirow}
\usepackage{algorithmic}
\usepackage{epstopdf}
\usepackage{color}
\usepackage{bigints}
\usepackage{mathtools}
\usepackage{relsize}

\usepackage[square, comma, sort&compress, numbers]{natbib}
\makeatletter
\newcommand{\vvast}{\bBigg@{3.0}}
\newcommand{\vast}{\bBigg@{4}}
\newcommand{\Vast}{\bBigg@{4.5}}
\newcommand{\VVast}{\bBigg@{5}}
\newcommand{\VVVast}{\bBigg@{5.5}}
\newcommand{\VVVVast}{\bBigg@{6}}

\makeatother

\newtheorem{theorem}{{Theorem}}

\newcommand{\ls}[1]
{\dimen0=\fontdimen6\the\font
	\lineskip=#1\dimen0
	\advance\lineskip.5\fontdimen5\the\font
	\advance\lineskip-\dimen0
	\lineskiplimit=.9\lineskip
	\baselineskip=\lineskip
	\advance\baselineskip\dimen0
	\normallineskip\lineskip
	\normallineskiplimit\lineskiplimit
	\normalbaselineskip\baselineskip
	\ignorespaces
}

\ifCLASSINFOpdf
\else
\fi
\IEEEoverridecommandlockouts\IEEEpubid{\makebox[\columnwidth]{979-8-3503-1090-0/23/\$31.00~\copyright~2023 IEEE \hfill} \hspace{\columnsep}\makebox[\columnwidth]{ }}

\begin{document}

	%\date{October 23, 2020} 
	
	\title{\ls{1.0}{Statistical AoI, Delay, and Error-Rate Bounded QoS Provisioning for Satellite-Terrestrial Integrated Networks}}

\pagenumbering{gobble}
\author{\IEEEauthorblockN{Jingqing Wang\IEEEauthorrefmark{2}, Wenchi Cheng\IEEEauthorrefmark{2}, and H. Vincent Poor\IEEEauthorrefmark{3}}~\\[0.2cm]
		\IEEEauthorblockA{\IEEEauthorrefmark{2}State Key Laboratory of Integrated Services Networks\\
			 Xidian University, Xi’an, China\\
			 	\IEEEauthorrefmark{3}Department of Electrical and Computer Engineering, Princeton University, Princeton, NJ 08544, USA\\
		E-mail: \{\emph{jqwangxd@xidian.edu.cn}, \emph{wccheng@xidian.edu.cn}, \emph{poor@princeton.edu}\}}

}
\maketitle

	\begin{abstract}
		Massive ultra-reliable and low latency communications (mURLLC) has emerged to support wireless time/error-sensitive services, which has attracted significant research attention while imposing several unprecedented challenges not encountered before.
	%	To upper-bound both delay and error-rate for mURLLC, statistical delay and error-rate bounded quality-of-services (QoS) provisioning has been developed as a powerful technique to characterize and implement QoS theory for timely and reliable wireless real-time traffic.
	%	Various 6G key enablers, such as statistical delay and error-rate bounded quality-of-services (QoS) provisioning, satellite communications, finite blocklength coding (FBC), etc., have been developed to support the demands of mURLLC.
		By leveraging the significant improvements in space-aerial-terrestrial resources for comprehensive 3D coverage, satellite-terrestrial integrated networks have been proposed to achieve rigorous and diverse quality-of-services (QoS) constraints of mURLLC.
		To effectively measure data freshness in satellite communications, recently, age of information (AoI) has surfaced as a novel QoS criterion for ensuring time-critical applications.
		Nevertheless, because of the complicated and dynamic nature of network environments, how to efficiently model multi-dimensional statistical QoS provisioning while upper-bounding peak AoI, delay, and error-rate for diverse network segments is still largely open. 
		To address these issues, in this paper we propose statistical QoS provisioning schemes over satellite-terrestrial integrated networks in the finite blocklength regime.   
		In particular, first we establish a satellite-terrestrial integrated wireless network architecture model and an AoI metric model. 
		Second, we derive a series of fundamental statistical QoS metrics including peak-AoI bounded QoS exponent, delay-bounded QoS exponent, and error-rate bounded QoS exponent. 
	Finally, we conduct a set of simulations to validate and evaluate our proposed statistical QoS provisioning schemes over satellite-terrestrial integrated networks.
	\end{abstract}
	
	%\vspace{-8pt}

\section{Introduction}\label{sec:intro}

\IEEEPARstart{W}{ith} 5G mobile wireless networks continuing to be globally commercialized, researchers are actively conceptualizing the upcoming 6G wireless networks~\cite{dang2020should} to ensure the overwhelmingly unparalleled modern wireless applications with increasingly rigorous and diverse  quality-of-services (QoS) constraints. 
As proposed in~\cite{cheng2015optimal}, the delay-bound QoS provisioning theory serves to portray the queueing dynamics that cater to the rising demands of wireless multimedia applications.
The inherently unstable nature of wireless fading channels and the intricate, diverse, and ever-changing structures of 6G mobile wireless networks make it extremely challenging to uphold the conventional deterministic QoS criteria for upcoming time- and error-sensitive multimedia data transmissions.
Towards this end, the \textit{statistical QoS provisioning theory} has emerged as a potential strategy for defining and implementing delay-bounded QoS guarantees for wireless multimedia communications.
However, the ever-increasing amount of time/error-sensitive multimedia traffic necessitates 6G wireless networks to meet increasingly complex and demanding QoS requirements, such as rigorous end-to-end delay, ultra-high reliability, unprecedented data rate, and extra-high energy efficiency, among others.
%6G networks are envisioned to support ultra-low latency, and massive connectivity for various emerging applications, including virtual reality, smart cities, autonomous vehicles, and beyond.

One of the key focuses of 6G development is to enable \textit{massive Ultra-Reliable Low-Latency Communications} (mURLLC)~\cite{9311792,HJ2018}, requiring stringent QoS guarantees to ensure their timely and reliable delivery over 6G mobile wireless networks. 
%Researchers have explored the implementation of small-packet communication techniques, such as finite blocklength coding (FBC)~\cite{yury2010,Yp2014}, as a means of reducing access latency and decoding complexity, while supporting the demanding QoS requirements of mission-critical mURLLC. The authors of~\cite{yury2010} have shown that the codeword blocklength can be as short as 100 channel symbols for the required reliable communications.  The maximum achievable coding rate using FBC over additive white Gaussian noise (AWGN) channels has been derived in~\cite{Yp2011}. The authors of~\cite{Yp2014} have studied different  properties of channel codes that approach the fundamental limits using FBC. The authors of~\cite{Mary2016} have exploited recent results on the non-asymptotic coding rate over time-varying wireless fading channel with no channel state information at transmitters (CSIT) to investigate the impact of the \textit{finite blocklength} on physical parameters of practical systems. 
However, one essential consideration in establishing comprehensive and various QoS standards for 6G mURLLC services is tackling the challenges imposed by massive access, specifically, those related to massive connectivity and massive coverage.
Towards this end, satellite communication systems have recently been developed to facilitate comprehensive 3D coverage for various 6G services involving space-aerial-terrestrial connectivity. 
These systems provide global coverage and ubiquitous connectivity while ensuring stringent QoS standards for mission-critical mURLLC.
On the other hand, conventional terrestrial networks still play an important role in providing low-cost and high-speed wireless services considering densely populated areas. 
Therefore, the combination of satellite and terrestrial networks has the potential in leveraging the advantages of both systems, enabling ubiquitous network service.

Furthermore, end-to-end satellite communication can reduce latency more effectively than terrestrial fiber optic networks.
Thus, satellite-terrestrial integrated wireless networks present significant potential in facilitating various emerging time-sensitive applications.
Given the frequent fluctuations in the status updates throughout satellite communications, it is imperative to prioritize the measurement and improvement of data freshness performance in support of real-time satellite telecommand/telemetry services, which is especially critical in the context of a vast amount of ground base stations (GBSs) necessary for mURLLC over 6G wireless networks.
The age of information (AoI)~\cite{9109636,yates2021age,7415972} represents a novel QoS benchmark to measure the degree of freshness with which a receiver can access information related to status updates at a remote data source. 
This metric is particularly important in satellite applications where delay- and age-sensitive data transmissions require accurate measurement of information freshness.
When contemplating AoI-driven QoS metrics over satellite-terrestrial integrated wireless networks, it is crucially important to consider that the status updates are usually made up of just a few information bits and must be transmitted to the remote destinations promptly. 
Therefore, it is imperative to consider small-packet communications, including finite blocklength coding (FBC)~\cite{yury2010,dosti2019performance}, as it holds significant importance over satellite-terrestrial integrated wireless networks.

Despite the diligent efforts in guaranteeing mURLLC via satellite-terrestrial integrated networks, a comprehensive wireless network architecture design paradigm that ensures stringent and diverse QoS provisioning has yet to be established.
Specifically, the lack of comprehensive understanding of QoS-driven fundamental performance-limits and modeling analyses, designing effective models for diverse QoS metrics while accounting for the diverse network segments remains a challenge.
Most existing works analyze system performances based on Shannon capacity with infinite blocklength.  
Considering statistical QoS theory, new analytical models are of paramount importance to support the rigorous QoS requirements of mURLLC over satellite-terrestrial integrated networks through FBC.

To solve the above challenges, in this paper we first propose the satellite-terrestrial integrated wireless networks using FBC and then develop a set of fundamental-performance metrics and their modeling techniques for statistical QoS provisioning.   
In particular, first we design the satellite-terrestrial integrated network architecture model and the AoI metric model.  
Second, focusing on modeling and analyzing the fundamental performance, we derive a number of novel statistical QoS metrics. 
Finally, we conduct a series of numerical results to verify, assess, and examine the developed statistical QoS performance modeling schemes over satellite-terrestrial integrated wireless networks using FBC.

The rest of this paper is organized as follows. Section~\ref{sec:sys} builds satellite-terrestrial integrated wireless system models. 
%Section~\ref{sec:pAoI} analytically characterizes the peak AoI violation probability functions. 
Section~\ref{sec:EC1} develops a set of new and fundamental statistical QoS metrics through FBC. 
Section~\ref{sec:results} verifies and examines our developed modeling schemes. The paper concludes with Section~\ref{sec:conclusion}.

	\section{The System Models}\label{sec:sys}
	\begin{figure}[!t]
	\vspace{-6pt}
	\centering
	\includegraphics[scale=0.37]{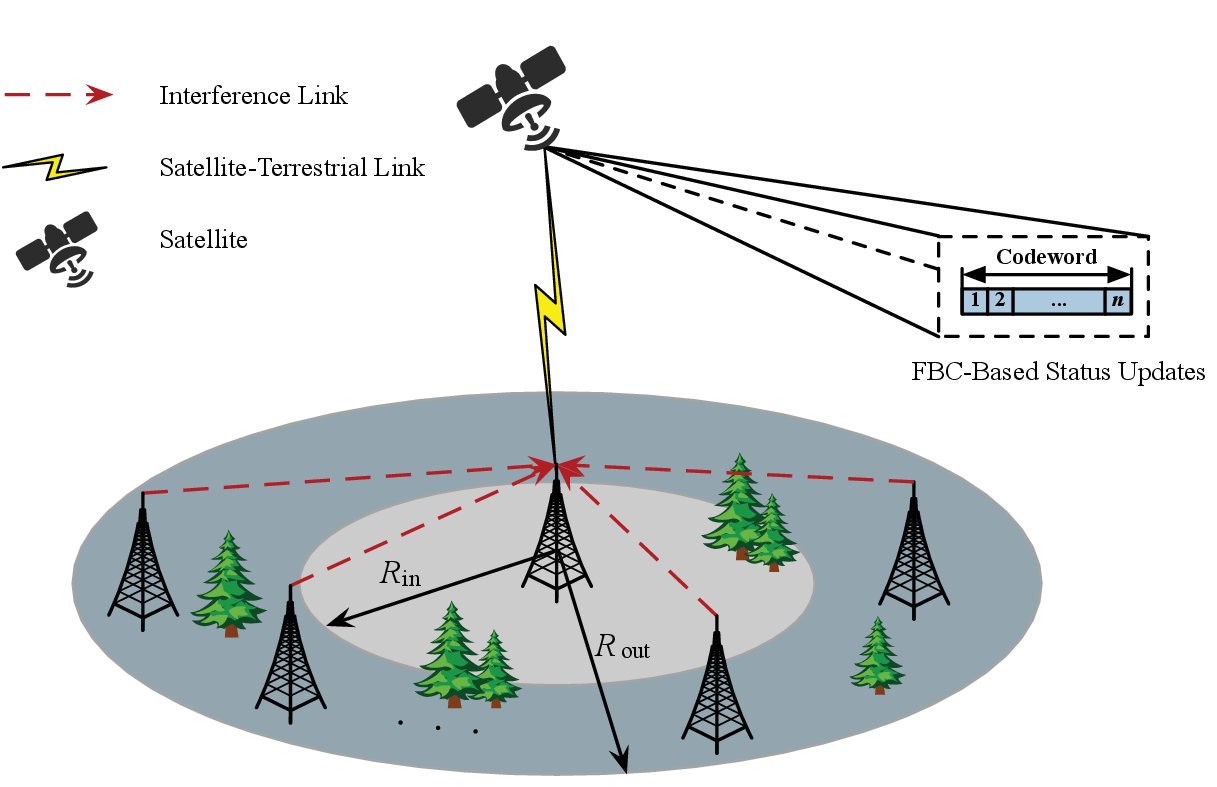}
	\caption{The AoI evolution for $N$ FBC-based status updates.}
	\label{fig:1}
	\vspace{-6pt}
\end{figure}
We consider the satellite-terrestrial integrated system architecture model, consisting of one satellite and $K$ GBSs, indexed by $\mathsf{K}=\{1,2,\dots, K\}$, which are randomly distributed around the destination GBS, as shown in Fig.~\ref{fig:1}.
We focus on the downlink transmission of remote sensing status updates, where satellite monitors dynamic status update packets and transmits the status updates to the destination GBS. % while guaranteeing statistical QoS requirements for mURLLC. 
We assume that each satellite is equipped with single antenna while each GBS is equipped with $N_{\text{R}}$ antennas.
We assume that $K$ nearby GBSs (terrestrial interferers) are located within a finite annular area centered around the destination node with an inner radius, denoted by $R_{\text{in}}$, and an outer radius, denoted by $R_{\text{out}}$, beyond which the interference is assumed to be negligible due to pathloss.
The proposed system model includes $N$ status-update data packets.
The index set, denoted by  $\textsf{N}$, represents all status-updates with a cardinality of $N$.

\subsection{The Wireless Channel Model Over Satellite-Terrestrial Integrated Networks Using FBC}

Define $h_{s}(u)$ $(u\in\textsf{N})$ as the channel fading coefficient between the satellite and destination node for transmitting status-update data packet $u$. We assume that $h_{s}(u)$ follows the shadowed-Rician distribution, which is commonly used to evaluate wireless land mobile satellite communication systems.
Given the short time span of transmitting data packets, it can be assumed that the wireless fading channel remains relatively stable during the codeword transmission.
Define the probability density function (PDF), denoted by $f_{|h_{s}(u)|^{2}}(x)$, of the channel gain $|h_{s}(u)|^{2}$ as follows:
\begin{equation}\label{equation_pdf}
	f_{|h_{s}(u)|^{2}}(x)=\alpha_{s}(u) e^{-\beta_{s}(u) x}  \sideset{_1}{_{1}}{\mathop{F}}(m_{s}(u),1,\delta_{s}(u) x), \,\, x>0, 
\end{equation}
where
\begin{equation}\label{equation02} 
	\begin{cases}
		\alpha_{s}(u)=\frac{1}{2b_{s}(u)}\left[\frac{2b_{s}(u)m_{s}(u)}{2b_{s}(u)m_{s}(u)+\Omega_{s}(u)}\right]^{m_{s}(u)};\\
		\beta_{s}(u) = \frac{1}{2b_{s}(u)};\\
		\delta_{s}(u)= \frac{\Omega_{s}(u)}{2b_{s}(u)\left[2b_{s}(u)m_{s}(u)+\Omega_{s}(u)\right]},
	\end{cases}
\end{equation}
where $\Omega_{s}(u)$ represents the average power of line-of-sight (LOS) component, $2b_{s}(u)$ is the average power for the multipath component, $m_{s}(u)\in[0,\infty]$ represents the Nakagami-$m$ parameter, and $_{1}F_{1}(\cdot,\cdot,\cdot)$ is the confluent hypergeometric function.
%When $m_{s}(u) = 0$, the shadowed-Rician fading reduces to Rayleigh fading. On the other hand, when $m_{s}(u) = \infty$, it converges to Rician fading.
Accordingly, the received signal vector, denoted by $\bm{y}_{s}(u)\in\mathbb{C}^{1\times n}$, from the satellite to the destination node for transmitting status update $u$ is derived as follows:
\begin{align}\label{equation03}
	\bm{y}_{s}(u)=&\sqrt{\left(\frac{c}{4\pi f_{c}d_{s}(u)}\right)^{2}G_{s}G_{d}(u)
		P_{s}(u)}h_{s}(u)\bm{x}_{s}(u)\nonumber\\
	&+\sum_{j=1}^{K}\sqrt{\left(\frac{c}{4\pi f_{c}d_{j}(u)}\right)^{2}G_{j}(u)G_{d}(u)P_{t}(u)}\nonumber\\
	&\times  g_{j}(u)\bm{x}_{j}(u) +\bm{n}_{s}(u)
\end{align}
where %$(\cdot)^{H}$ is the Hermitian transpose of a matrix,
$\bm{x}_{s}(u)$ and $\bm{x}_{j}(u)$ are the data streams from the satellite and GBS $j$, respectively, $c$ denotes the speed of light, $f_{c}$ is the frequency, $d_{s}(u)$ and $d_{j}(u)$ denote the distances between the satellite and destination node and between GBS $j$ and the destination node, respectively, $G_{s}$ is the antenna gain at the satellite, $G_{d}(u)$ and $G_{j}(u)$ are the antenna gains at the destination node and GBS $j$, respectively,  
$g_{j}$ is the terrestrial channel fading coefficient from the GBS $j$ to the destination node, which follows Rayleigh distribution, $P_{s}(u)$ and $P_{t}(u)$ are the transmit powers at the satellite and GBS $j$, respectively, and $\bm{n}_{s}(u)\sim{\cal CN}(0,\sigma^{2})$ is the additive white Gaussian noise (AWGN) vector.
Then, the signal-to-interference-plus-noise ratio (SINR), denoted by $\gamma_{s}(u)$, is derived as follows: 
\begin{equation}\label{equation04}
	\gamma_{s}(u)=\frac{\phi_{s}(u){\cal P}_{s}(u)|h_{s}(u)|^{2}}{I_{\text{a}}+1}
\end{equation}
where ${\cal P}_{s}(u)=P_{s}(u)/\sigma^{2}$ represents the transmit signal-to-noise ratio (SNR), and 
\begin{equation}\label{equation05}
	\begin{cases}
			\phi_{s}(u)\triangleq\left(\frac{c}{4\pi f_{c}d_{s}(u)}\right)^{2}G_{s}G_{d}(u);\\
			\phi_{j}(u)\triangleq\left(\frac{c}{4\pi f_{c}d_{j}(u)}\right)^{2}G_{j}(u)G_{d}(u), \quad j\in\mathsf{K},
	\end{cases}
\end{equation}
and $I_{\text{a}}$ is the aggregate interference power received from terrestrial interferers (nearby GBSs) with the transmit power, denoted by $P_{t}(u)$, which is obtained as follows:
\begin{equation}
	I_{\text{a}}=\sum\limits_{j=1}^{K}\phi_{j}(u)P_{t}(u)|h_{j}(u)|^{2}
\end{equation}
where ${\cal P}_{t}(u)={\cal P}_{t}(u)/\sigma^{2}$ represents the transmit SNR of the terrestrial interferers. 

Moreover, to support \textit{statistical} QoS provisioning through FBC, we derive the decoding error probability over shadowed-Rician wireless fading channels.
The average decoding error probability, denoted by $\epsilon_{s}(u)$, from the satellite to destination node when transmitting status update $u$ is given as follows:
\begin{equation}\label{equation09}
	\epsilon_{s}(u)\approx \mathbb{E}_{\gamma_{s}(u)} \left[{\cal Q}\left(\frac{\left(C(\gamma_{s}(u))-R^{*}_{s}\right)}{\sqrt{V(\gamma_{s}(u))/n}}\right)\right]
\end{equation}
where $\mathbb{E}_{\gamma_{s}(u)}[\cdot]$ is the expectation over the SINR $\gamma_{s}(u)$, ${\cal Q}(x)=1/\sqrt{2\pi}\int_{x}^{\infty}\exp(-t^{2}/2)dt$, $R^{*}_{s}$ is the maximum achievable coding rate, and $C(\gamma_{s}(u))$ and $V(\gamma_{s}(u))$ are the \textit{channel capacity} and \textit{channel dispersion}, respectively.

\subsection{The AoI Metric Modeling Through FBC}
	\begin{figure}[!t]
		\vspace{-6pt}
	\centering
	\includegraphics[scale=0.31]{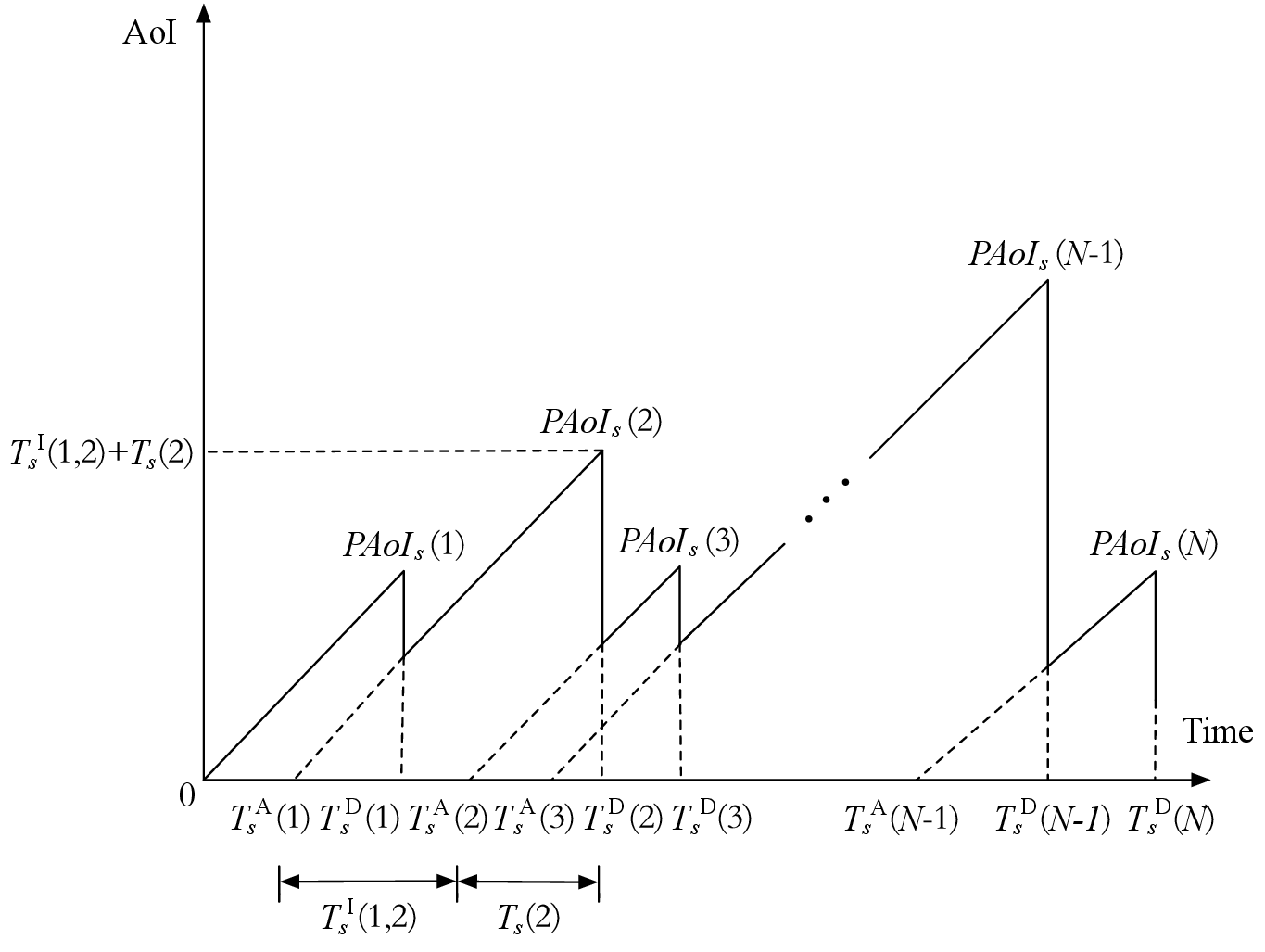}
	\caption{The AoI evolution for $N$ FBC-based status updates.}
	\label{fig:2}
	\vspace{-6pt}
\end{figure}

To quantify/manage the timeliness of information, we employ the AoI metric as a fundamental QoS measure to indicate the recency of received data.
As illustrated in Fig.~\ref{fig:2}, for $u\in\textsf{N}$, we define $T_{s}^{\text{A}}(u)$, $T_{s}^{\text{D}}(u)$, and $T_{s}^{\text{S}}(u)$ as the arrival, departure, and service times for transmitting FBC-based status update $u$, respectively. 
In addition,  $T_{s}^{\text{A}}(0)$ is set to be zero.	
Denote by $T_{s}^{\text{I}}(v,u)$ the time of inter-arrival for transmitting between status update $v$ and status update $u$ $(1\leq v\leq u)$, as shown in Fig.~\ref{fig:2}, where we have $T_{s}^{\text{I}}(v,u)\triangleq T_{s}^{\text{A}}(u)-T_{s}^{\text{A}}(v)$. 
%Then, denote by $T_{s}^{\text{I}}(u-1,u)$ the inter-arrival time between the $(u-1)$th and the $u$th status-update data packets. 
We can derive the inter-arrival time $T_{s}^{\text{I}}(v,u)$ as in the following equation:
\begin{equation}\label{equation003}
	T_{s}^{\text{I}}(v,u)=\sum_{i=v+1}^{u}T_{s}^{\text{I}}(i-1,i).
\end{equation}
Then, we can derive the time for cumulative service, denoted by $T_{s}^{\text{S}}(v,u)$, when transmitting status update $v$ to status update $u$ as follows:
\begin{equation}
	T_{s}^{\text{S}}(v,u)=\sum_{i=v}^{u}T_{s}^{\text{S}}(i).
\end{equation}
Correspondingly, the time of departure $T_{s}^{\text{D}}(u)$ for transmitting $u$ ($u\geq 1$)  update data packets is derived as follows:
\begin{align}\label{equation015}
	T_{s}^{\text{D}}(u)=\max_{v\leq u}\left\{T_{s}^{\text{A}}(v)+T_{s}^{\text{S}}(v,u)\right\}.
\end{align}
Accordingly, the total sojourn time, denoted by $T_{s}(u)$, when transmitting status update $u$ is determined as follows:
\begin{equation}\label{equation005}
	T_{s}(u)%= T_{s}^{\text{D}}(u)-T_{s}^{\text{A}}(u)
	=\max_{v\in\textsf{N}, v\leq u}\left\{T_{s}^{\text{S}}(v,u)-T_{s}^{\text{I}}(v,u)\right\}.
\end{equation}
Fig.~\ref{fig:2} shows that the peak AoI, denoted by $PAoI_{s}(u)$, for our proposed satellite-terrestrial integrated system model when transmitting status-update data packet $u$ is expressed as in the following equation:
%\begin{equation}\label{equation06}
%	PAoI_{s}(u)
%	=T_{s}^{\text{D}}(u)-T_{s}^{\text{A}}(u-1).
%\end{equation}
%Then, using Eqs.~(\ref{equation003}),~(\ref{equation005}),  and~(\ref{equation06}), the peak AoI $PAoI_{s}(u)$  for the $u$th status-update data packet can be rewritten as follows:
\begin{equation}\label{equation08}
	PAoI_{s}(u)=T_{s}^{\text{I}}(u-1,u)+T_{s}(u).
\end{equation}
%	Assume that the processing delay and the propagation delay are negligible. Let 
%	\begin{equation}
	%	T_{s}(u)=W(u)+T_{s}^{\text{S}}(u)
	%	\end{equation}
%	where $W(u)$ is the waiting time in the queue for the $u$th status-update data packet and $T_{s}^{\text{S}}(u)$ is the service time of the $u$th status-update data packet.

\section{Statistical QoS Metrics Over Satellite-Terrestrial Integrated Networks}\label{sec:EC1}

One critical aspect to consider for the design of satellite-terrestrial integrated wireless services is the efficient guarantee of low-tail requirements for age/time/reliability-sensitive data transmissions. 
Rather than solely evaluating performance metrics on average, our proposed schemes must prioritize supporting QoS provisioning for lower-tail cases, which entails focusing on assessing the probability of service constraints being violated. %in order to ensure optimal QoS provisioning.
Consequently, this section aims to characterize the violation probabilities with regard to peak AoI, delay, and error-rate through FBC. This aspect plays a significant role in our fundamental performance analyses.

\subsection{The Peak-AoI Bounded QoS Metrics Using FBC}\label{sec:pAoI}

To support the peak-AoI bounded QoS provisioning, we propose to conduct a characterization of the probability of peak AoI violation, which is determined by the probability that the peak AoI exceeds the designated peak AoI threshold.

\textit{Definition 1: The peak-AoI bounded QoS exponent:} Using the \textit{large deviations principle} (LDP), the peak-AoI bounded QoS exponent, denoted by $\theta_{\text{AoI}}$, characterizes the relationship between the peak AoI threshold and the probability of peak AoI exceeding the given threshold, which measures the exponentially decaying rate of the peak AoI violation probability. 	

Note that the peak-AoI bounded QoS exponent $\theta_{\text{AoI}}$ is an indicator of the level of stringency associated with statistical peak-AoI bounded QoS provisioning as the threshold increases. 
Accordingly, our proposed strategy aims to alleviate the tail behaviors regarding the peak AoI metric.
%The probability of violating the peak AoI threshold is denoted by the peak AoI violation probability for a given designated peak AoI threshold. 
This peak-AoI violation probability quantitatively represents the extreme behaviors of the peak AoI, which can then be used to compare with the delay and error-rate violation probabilities.

In order to ensure compliance with the demanding mURLLC standards, it is imperative to incorporate peak AoI violation probability as a novel QoS standard. This facilitates the systematic evaluation, administration, and assessment of the age/delay process for effective statistical QoS provisioning.
Due to the stochastic nature of arrival and service processes, it is not feasible to determine the peak AoI violation probability directly. 
To address this, we propose to implement stochastic network calculus (SNC), which employs Mellin transform to transform random arrival/service time in the exponential domain.
By converting the inter-arrival time and service time into the exponential domain, we have 
	\begin{equation}
		\begin{cases}
			{\cal T}_{s}^{\text{I}}(v,u)\triangleq e^{T_{s}^{\text{I}}(v,u)}; \\ 
			{\cal T}_{s}^{\text{S}}(v,u)\triangleq e^{T_{s}^{\text{S}}(v,u)}.
		\end{cases}
	\end{equation}
	Based on Eq.~(\ref{equation08}), we can derive the peak AoI metric in the exponential domain, denoted by $\mathsf{PAoI}_{s}(u)$, from the satellite to the destination node for transmitting status-update $u$ as follows:
	\begin{align}\label{equation11}
		\mathsf{PAoI}_{s}(u)&=e^{PAoI_{s}(u)}
		={\cal T}_{s}^{\text{I}}(u-1,u){\cal T}_{s}(u)
	\end{align}
	where ${\cal T}_{s}^{\text{I}}(u-1,u)$ and ${\cal T}_{s}(u)$ are the time of inter-arrival between status update $(u-1)$ and status update $u$ and total sojourn time in the exponential domain, respectively, which are determined as follows:
	\begin{equation}\label{equation12}
		\begin{cases}
			{\cal T}_{s}^{\text{I}}(u-1,u)=e^{T_{s}^{\text{I}}(u-1,u)};\\ %=\frac{{\cal A}(u)}{{\cal A}(u-1)};\\
			{\cal T}_{s}(u)=e^{ T_{s}(u)}. %=\frac{{\cal D}(u)}{{\cal A}(u)}.
		\end{cases}
	\end{equation}
	Then, we can derive the peak AoI violation probability, denoted by $p_{\text{PAoI},s}(u)$, for transmitting status-update $u$ as follows:
	\begin{align}\label{equation49}
		%\lim\limits_{t^{(n)}\rightarrow \infty}
		p_{\text{PAoI},s}(u)\triangleq\text{Pr}\left\{ PAoI_{s}(u)>\frac{A_{\text{th}}}{n}\right\}
		\leq e^{-\frac{A_{\text{th}}}{n}\theta_{\text{AoI}}}
	\end{align}
where $A_{\text{th}}$ is the threshold for peak AoI in the number of channel uses through using FBC.
Since it is not feasible to directly compute the exact peak AoI violation probability, an upper-bound estimation can be obtained by utilizing the Mellin transform.
%Define the Mellin transform of a non-negative random variable ${\cal X}$ as ${\cal M}_{{\cal X}}(\theta)\triangleq \mathbb{E}\left[{\cal X}^{(\theta-1)}\right]$, 
	%\begin{equation}\label{equation10}
	%	{\cal M}_{{\cal X}}(\theta)\triangleq \mathbb{E}\left[{\cal X}^{(\theta-1)}\right]
	%\end{equation}
	%Considering a G/G/1 queue, 
Then, the Mellin transform in terms of the peak AoI, denoted by ${\cal M}_{\mathsf{PAoI}_{s}(u)}(\theta_{\text{AoI}})$, is derived as follows:
	\begin{align}
		{\cal M}_{\mathsf{PAoI}_{s}(u)}(\theta_{\text{AoI}})&=\mathbb{E}\left[\left(\mathsf{PAoI}_{s}(u)\right)^{(\theta_{\text{AoI}}-1)}\right]
	\nonumber\\
	&={\cal M}_{{\cal T}_{s}^{\text{I}}(u-1,u)}(\theta_{\text{AoI}}){\cal M}_{{\cal T}_{s}(u)}(\theta_{\text{AoI}})
	\end{align}
	where $\mathbb{E}[\cdot]$ represents the expectation operation, 
	${\cal M}_{{\cal T}_{s}^{\text{I}}(u-1,u)}(\theta_{\text{AoI}})$ and ${\cal M}_{{\cal T}_{s}(u)}(\theta_{\text{AoI}})$ are the Mellin transforms of the inter-arrival time and sojourn time, respectively.
	Based on the Mellin transform, we present a thorough derivation of the peak AoI violation probability, as stated in the following theorem.

	\begin{theorem}\label{theorem1}
		The upper-bounded peak AoI violation probability $p_{\text{PAoI},s}(u)$ for our proposed satellite-terrestrial integrated wireless networks for a given the peak AoI threshold $A_{\text{th}}$ is determined as follows:
		\begin{align}\label{equation026a}
			p_{\text{PAoI},s}(u)&\leq e^{- \frac{\theta_{\text{AoI}} A_{\text{th}}}{n}}\mathsf{K}(\theta_{\text{AoI}},u)
		\end{align}
		where
		\begin{align}\label{equation027a}
			\mathsf{K}_{s}(\theta_{\text{AoI}},u)\triangleq&{\cal M}_{{\cal T}_{s}^{\text{I}}(u-1,u)}(1+\theta_{\text{AoI}})
			\Bigg[\sum_{v=1}^{u} {\cal M}_{{\cal T}_{s}^{\text{S}}(v,u)}(1+\theta_{\text{AoI}})
			\nonumber\\
			&\times 
		 {\cal M}_{{\cal T}_{s}^{\text{I}}(v,u)}(1-\theta_{\text{AoI}})\Bigg].
		\end{align}
	\end{theorem}
	
	\begin{IEEEproof}
		The proof of Theorem~\ref{theorem1} is omitted in this paper due to lack of space.
	\end{IEEEproof}

\subsection{The Delay-Bounded QoS Metrics}
Research has been conducted on statistical delay-bounded QoS guarantees to investigate the queuing dynamics in the presence of arrival and service processes that vary over time.

\textit{Definition 2: The statistical delay-bounded QoS exponent:} By using the LDP, the queueing process converges to a random variable $Q_{s}(\infty)$ in distribution for satellite-terrestrial integrated wireless communications such that
\begin{equation}\label{equation23}
	-\lim_{Q_{\text{th},s}\rightarrow\infty}\frac{\log\left(\text{Pr}
		\left\{Q_{s}(\infty)>Q_{\text{th},s}\right\}\right)}{Q_{\text{th},s}}=\theta_{\text{delay}}
\end{equation}
where $Q_{\text{th},s}$ denotes the threshold of buffer overflow and $\theta_{\text{delay}}$ $(\theta_{\text{delay}}>0)$ represents the delay-bounded QoS exponent to describe the queuing delay, where it quantifies the rate of exponential decay that occurs in the delay-bounded QoS violation probabilities. 		
Specifically, a sufficiently small delay-bounded QoS exponent enables the satellite-terrestrial integrated system to accommodate an arbitrarily prolonged delay, whereas a sufficiently large delay-bounded QoS exponent renders the system intolerant of any delay.

Furthermore, the Mellin transform in terms of the service process, denoted by ${\cal M}_{{\cal S}_{s}(u)}(\theta_{\text{delay}})$, can be derived as follows:
\begin{align}\label{equation83}
	{\cal M}_{{\cal S}_{s}(u)}(\theta_{\text{delay}})
	&=\mathbb{E}_{\gamma_{s}(u)}\left[\epsilon\left(\gamma_{s}(u)\right)\right] +\mathbb{E}_{\gamma_{s}(u)}\left[1-\epsilon\left(\!\gamma_{s}(u)\right)\right] 
		\nonumber\\
	&\quad\times e^{(\theta_{\text{delay}}-1)  \log_{2}M_{s}}.
\end{align}
where $M_{s}$ is the code size.
Define ${\cal M}_{{\cal A}_{s}}(\theta_{\text{delay}})$ as the Mellin transform in terms of the arrival process, denoted by $A_{s}(u)$.
Given the target delay $D_{\text{th}}$, a kernel function $\widetilde{\mathsf{K}}_{s}(\theta_{\text{delay}},u)$ for measuring the queuing delay is defined as follows:
\begin{equation}\label{equation1126}
	\widetilde{\mathsf{K}}_{s}(\theta_{\text{delay}},u)\!\triangleq\! \frac{\left[{\cal M}_{{\cal S}_{s}(u)}(1-\theta_{\text{delay}})\right]^{D_{\text{th}}}}{1\!-\!{\cal M}_{{\cal A}_{s}}\!(\theta_{\text{delay}})(1\!+\!\theta_{\text{delay}}){\cal M}_{{\cal S}_{s}\!(u)}(1\!-\!\theta_{\text{delay}})},
\end{equation}
if the following stability condition can hold:
\begin{equation}
	{\cal M}_{{\cal A}_{s}}(\theta_{\text{delay}})(1+\theta_{\text{delay}}){\cal M}_{{\cal S}_{s}(u)}(1-\theta_{\text{delay}})<1.
\end{equation}
As a result, through implementing the Mellin transform for the arrival and service processes $A_{s}(u)$ and $S_{s}(u)$ in the exponential domain, we can derive an upper-bounded delay violation probability, denoted by $p_{\text{delay},s}(u)$, as follows:
\begin{equation}\label{equation127}
	p_{\text{delay},s}(u)\leq  \inf_{\theta_{\text{delay}}>0}\left\{\widetilde{\mathsf{K}}_{s}(\theta_{\text{delay}},u)\right\}.
\end{equation}

\subsection{The Error-Rate Bounded QoS Metrics}

Previous research work has been conducted on the tradeoff between reliability and data transmission rate in the realm of small-packet communications by characterizing the \textit{error exponents}.  
Consequently, to represent the error-rate bounded QoS metrics in the context of satellite-terrestrial integrated wireless networks, it is crucial to explore the connections between the \textit{error-rate bounded QoS exponent} and the maximum feasible coding rate.

\textit{Definition 3: The error-rate bounded QoS exponent:} By using the LDP, the error-rate bounded QoS exponent, denoted by $\theta_{\text{error}}$, characterizes the \textit{exponential decaying rate} of the decoding error probability $\epsilon_{s}(u)$, which is defined as follows:
\begin{equation}\label{equation38}
	\theta_{\text{error}}\triangleq\lim\limits_{n\rightarrow \infty}-\frac{1}{n}\log(\epsilon_{s}(u))
\end{equation}
when the coding rate falls below the channel capacity, i.e. $R^{*}_{s}<C(\gamma_{s}(u))$.
Note that Eq.~(\ref{equation38}) indicates that the decoding error probability $\epsilon_{s}(u)$ decays exponentially at the rate of $\theta_{\text{error}}$, measuring the \textit{stringency} of statistical error-rate bounded QoS, as the blocklength $n$ increases.
Correspondingly, for a given coding rate $R^{*}_{s}<C(\gamma_{s}(u))$, the decoding error probability $\epsilon_{s}(u)$ vanishes exponentially as $n\rightarrow\infty$, i.e.,
\begin{equation}\label{equation039}
	\epsilon_{s}(u)\leq\exp\left[-n\theta_{\text{error}}\right]=\exp(-n\theta_{\text{error}}).
\end{equation}
Note that Eq.~\eqref{equation039} shows the relationship among the decoding error probability $\epsilon_{s}(u)$, blocklength $n$, and the \textit{maximum achievable coding rate} $R^{*}_{s}$. 
Based on the definition in~\cite{gallager1968information}, we can obtain the \textit{error-rate bounded QoS exponent} $\theta_{\text{error}}$ as follows:
\begin{equation}\label{equation040}
	\theta_{\text{error}}\triangleq\sup_{ \rho_{s}\in[0,1]}\left\{E_{0}\left[\rho_{s},P_{\bm{x}_{s}}\left(\bm{x}_{s}\right)\right]-\rho_{s} R^{*}_{s}\right\}
\end{equation}
where $\rho_{s}$ represents the Lagrange multiplier parameter, which is a real-valued number satisfying $\rho_{s}\in[0,1]$, and $P_{\bm{x}_{s}}\left(\bm{x}_{s}\right)$ is the PDF of the transmitted signal vector $\bm{x}_{s}$ from the satellite to the destination node, and
\begin{align}\label{equation041}
	&E_{0}\left[\rho_{s},P_{\bm{x}_{s}}\left(\bm{x}_{s}\right)\right]\triangleq\! -\frac{1}{n}\log\! \Bigg\{\!\mathbb{E}_{\gamma_{s}(u)}\!\Bigg[\int_{\bm{y}_{s}}\!\Bigg[\int_{\bm{x}_{s}}\!P_{\bm{x}_{s}}
	\nonumber\\
	&\qquad\times \left(\bm{x}_{s}\right)\!\left[P_{\bm{y}_{s}|\bm{x}_{s},h_{s}}\!\left(\bm{y}_{s}|\bm{x}_{s},h_{s}\right)\right]^{\frac{1}{(1+\rho_{s})}}d\bm{x}_{s}\Bigg]^{(1+\rho_{s})}\!\!\!d\bm{y}_{s}\!\Bigg]\!\Bigg\}
	\nonumber\\
	&\quad=-\frac{1}{n}\log \left\{\mathbb{E}_{\gamma_{s}(u)}\left[\left(1+\frac{\gamma_{s}(u)}{1+\rho_{s}}\right)^{-n\rho_{s}}\right]\right\}
\end{align}
where $P_{\mathbf{Y}|\bm{x}_{s},h_{s}}\left(\mathbf{Y}|\bm{x}_{s},h_{s}\right)$ is the conditional PDF.
Then, we can derive the error-rate bounded QoS exponent $\theta_{\text{error}}$ from the satellite to the destination node for our proposed performance modeling schemes as follows:
\begin{align}\label{equation046}
	\theta_{\text{error}}=&\sup_{ \rho_{s}\in[0,1]}\Bigg\{-\frac{1}{n}\log \left\{\mathbb{E}_{\gamma_{s}(u)}\left[\left(1+\frac{\gamma_{s}(u)}{1+\rho_{s}}\right)^{-n\rho_{s}}\right]\right\}
	\nonumber\\
	&\qquad\qquad-\rho_{s} R^{*}_{s}\Bigg\}.
\end{align}

The most challenging part in deriving the closed-form expression for the \textit{error-rate bounded QoS exponent} lies in obtaining the closed-form expression of $E_{0}\left[\rho_{s},P_{\bm{x}_{s}}\left(\bm{x}_{s}\right)\right]$.
Defining $\rho_{s}^{*}$ as the optimal Lagrange multiplier parameter that maximizes the error-rate bounded QoS exponent $\theta_{\text{error}}$, we can investigate the error-rate bounded QoS exponent as in the following theorem.

\begin{theorem}\label{theorem05}
	The error-rate bounded QoS exponent $\theta_{\text{error}}$ for the FBC-based wireless transmissions between the satellite and the destination node is approximately determined as follows: 
	\begin{align}\label{theorem05_eq00}
		\theta_{\text{error}}
		\approx &\!\left[\log\!\left(\frac{2{\cal P}_{s}(u)N_{\text{R}}\!+\!2N_{\text{R}}\sum\limits_{j=1}^{K}{\cal P}_{t}(u)\!+\!1}{2N_{\text{R}}\sum\limits_{j=1}^{K}{\cal P}_{t}(u)+1}\right)\!-\!R^{*}_{s}\right]^{2}
		\nonumber\\
		&\times 
		\left[4-\frac{2(2N_{\text{R}}\sum\limits_{j=1}^{K}{\cal P}_{t}(u)+1)}{2{\cal P}_{s}(u)N_{\text{R}}+2N_{\text{R}}\sum\limits_{j=1}^{K}{\cal P}_{t}(u)+1}\right]^{-1}.
	\end{align}
\end{theorem}

\begin{IEEEproof}
The proof of Theorem~\ref{theorem05} is omitted in this paper due to lack of space.
\end{IEEEproof}

%\indent\textit{Remarks on Theorem~\ref{theorem05}:} Theorem~\ref{theorem05} investigates the \textit{fundamental-performance} while focusing on investigating the error-rate bounded QoS exponent that is the rate at which the error probability decays exponentially with blocklength $n$ when blocklength $n\rightarrow\infty$. Theorem~\ref{theorem05} plays a critically important role in characterizing the statistical error-rate bounded QoS provisioning.

\section{Performance Evaluations}\label{sec:results}

We provide a set of numerical results to verify and analyze our proposed statistical QoS provisioning schemes over satellite-terrestrial integrated mobile wireless networks. 
%Consider that GBSs are randomly and uniformly distributed in a region with a radius of 2 km.
Before the simulation, it is necessary to determine the relationship between channel use (cu) and second. 
Assuming that 2PSK modulation is employed and the transmission rate of the satellites is 1Mbps, then we obtain 1 cu = $10^{-6}$ s. 
In our simulation scenario, we set the outer radius $R_{\text{out}}=10$ Km, inner radius $R_{\text{in}}=2$ Km,  the antenna gain at satellite $G_{s}=20$ dBi, and the transmit power at the satellite $P_{s}(u)\in[10,50]$ dBm.

\begin{figure}[!t]
	\vspace{-6pt}
	\centering
	\includegraphics[scale=0.3]{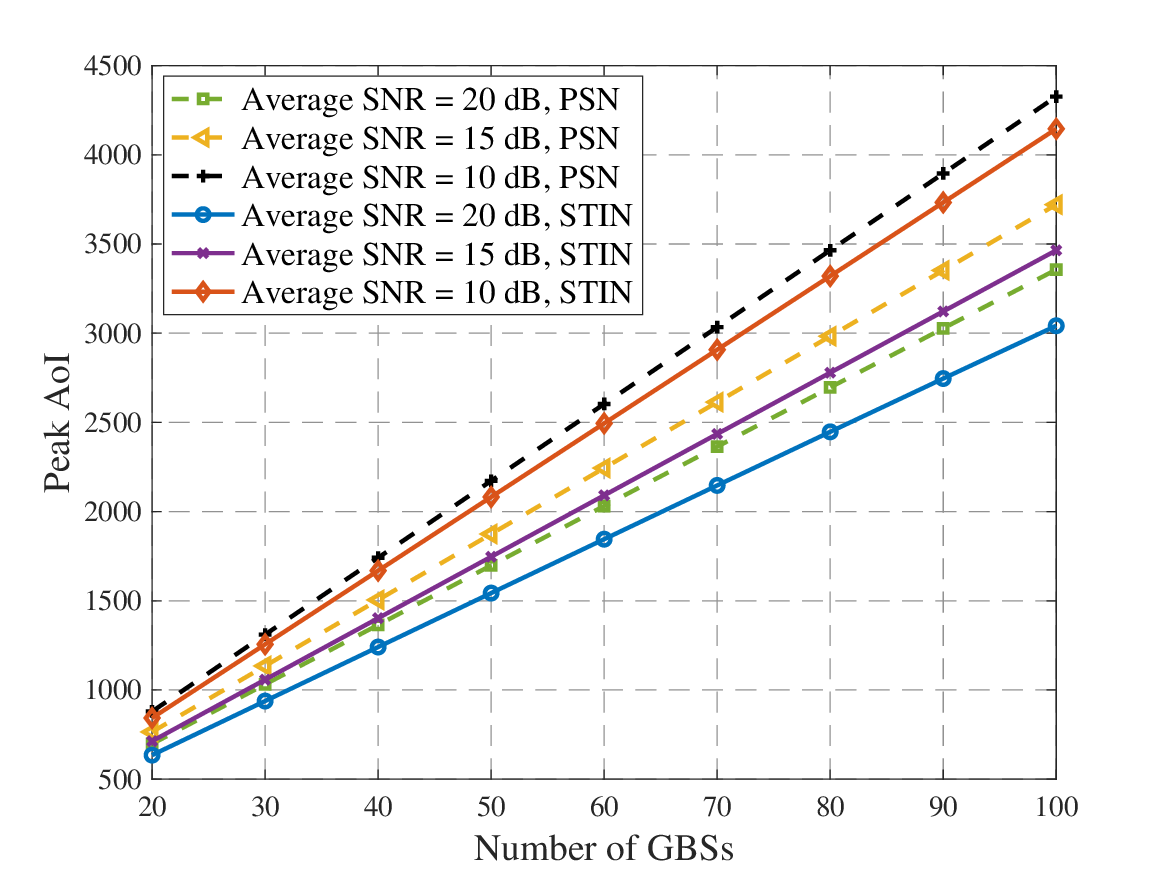}
	\caption{The peak AoI (cu) vs. the number of GBSs $K$ through using FBC.}
	\label{fig:03}
	\vspace{-6pt}
\end{figure}

Figure~\ref{fig:03} plots the peak AoI in terms of channel uses against the number of GBSs over the developed satellite-terrestrial integrated networks (STIN) as compared with the pure satellite networks (PSN). 
Fig.~\ref{fig:03} shows the peak AoI metric increases with more deployed GBSs. Since an increased number of GBSs in each retransmission cycle occupy more time slots, leading to a subsequent rise in the queuing delay.
Fig.~\ref{fig:03} also shows that the proposed satellite-terrestrial integrated schemes achieve better peak AoI performance when compared to the PSN since the peak AoI in terrestrial systems is relatively smaller than that of the pure satellite systems. 
Moreover, Fig.~\ref{fig:03} illustrates that the peak AoI decreases as the average SNR increases. %This implies that more satellites provide additional access points for the GBSs, which results in the reduction of the queueing time for uploading data packets and consequently reduces the peak AoI.

\begin{figure}[!t]
	\vspace{-6pt}
	\centering
	\includegraphics[scale=0.3]{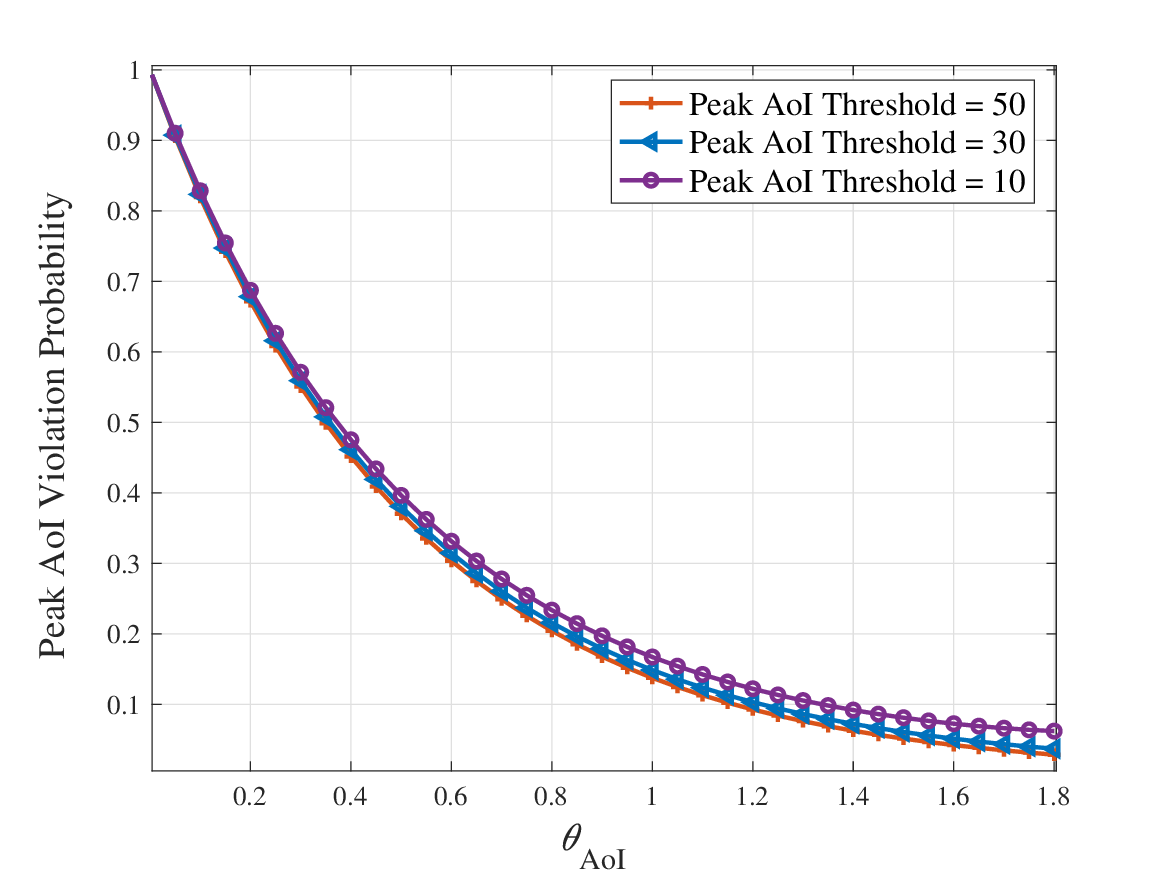}
	\caption{The peak AoI violation probability vs. the peak-AoI bounded QoS exponent $\theta_{\text{AoI}}$ for our developed satellite-terrestrial integrated networks.}
	\label{fig:04}
	\vspace{-6pt}
\end{figure}

\begin{figure}[!t]
	\vspace{-6pt}
	\centering
	\includegraphics[scale=0.3]{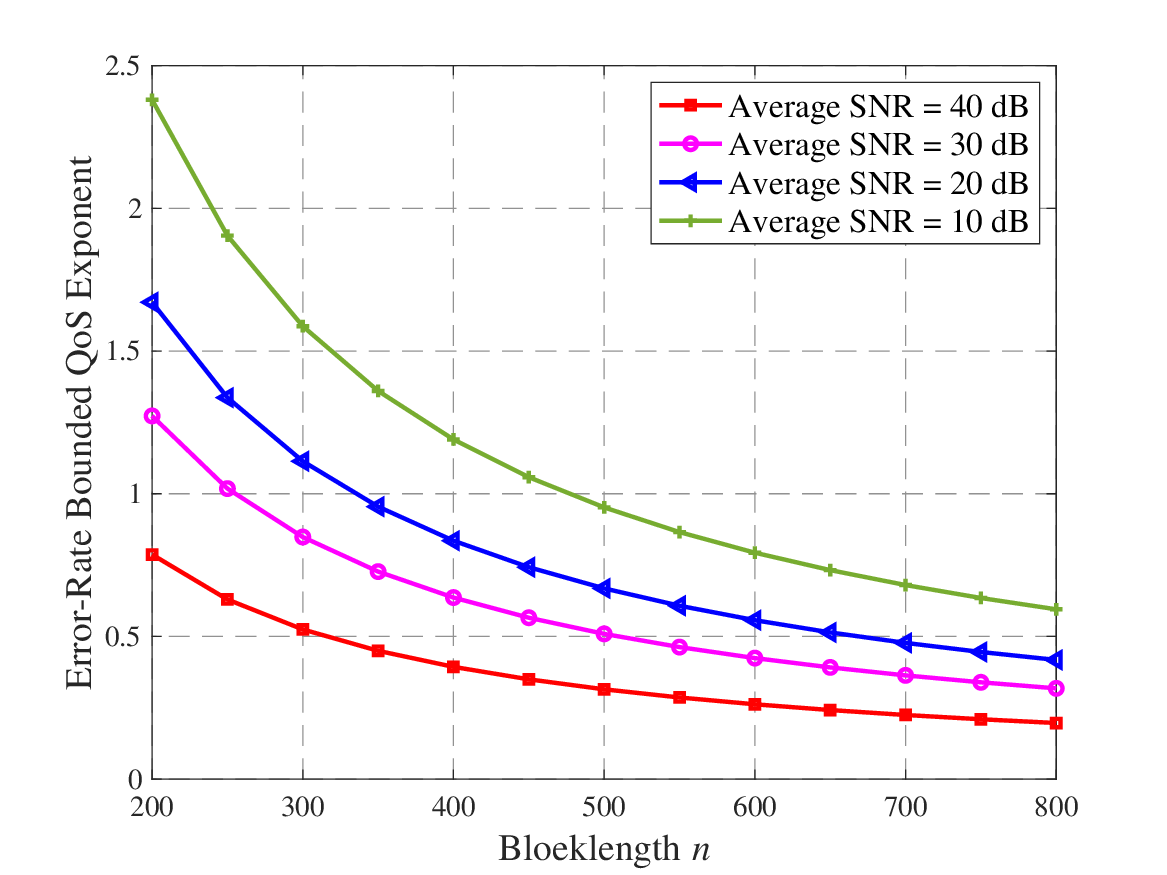}
	\caption{The error-rate bounded QoS exponent $\theta_{\text{error}}$ vs. the blocklength $n$.}
	\label{fig:05}
	\vspace{-6pt}
\end{figure}

Figure~\ref{fig:04} plots the peak AoI violation probability against the peak-AoI bounded QoS exponent $\theta_{\text{AoI}}$. 
Fig.~\ref{fig:04} demonstrates that the peak AoI violation probability decreases as the peak-AoI bounded QoS exponent increases, which validates the definition of $\theta_{\text{AoI}}$.
The peak AoI violation probability is bounded between a lower limit set by a large peak-AoI bounded QoS exponent and an upper limit set by a small peak-AoI bounded QoS exponent.
In addition, Fig.~\ref{fig:05} plots the error-rate bounded QoS exponent against the length of the codeword $n$. 
Fig.~\ref{fig:05} shows that the error-rate bounded QoS exponent decays exponentially with blocklength $n$ when blocklength $n\rightarrow\infty$, validating the definition of the error-rate bounded QoS exponent $\theta_{\text{error}}$.

	\section{Conclusions}\label{sec:conclusion} 

We have proposed a satellite-terrestrial integrated mobile wireless network architecture and developed a series of fundamental-performance metrics and their modeling techniques for statistical QoS provisioning.   
Particularly, first we have developed satellite-terrestrial integrated wireless network models.
Second, we have developed a number of new fundamental statistical-QoS performance metrics using FBC. 
Finally, a set of simulations have been conducted to verify, assess, and examine our established statistical QoS provisioning schemes for satellite-terrestrial integrated 6G mobile wireless networks.

\section*{Acknowledgment}
This work was supported in part by the National Key R\&D Program of China under Grant 2021YFC3002102 and in part by the Key R\&D Plan of Shaanxi Province under Grant 2022ZDLGY05-09, and in part by the U.S National Science Foundation under Grant CNS-2128448.

	\nocite{*}
\footnotesize
\bibliographystyle{IEEEtran}
\bibliography{myref.bib}
%\bibliography{IEEEabrv, myref}

\end{document}